\documentclass[useAMS,usenatbib]{mn2e}
\usepackage[]{graphicx, natbib, lscape}

\def\be{\begin{equation}}
\def\ee{\end{equation}}
\def\bea{\begin{eqnarray}}
\def\eea{\end{eqnarray}}
\newcommand{\rmd}{{\rm{d}}}

\title[The impact of camera optical alignments on weak lensing measures for the Dark Energy Survey]
{The impact of camera optical alignments on weak lensing measures for the Dark Energy Survey}
\author[M. Antonik, D. J. Bacon, S. L. Bridle, P. Doel et al.]{Michelle L. Antonik$^1$, David J. Bacon$^2$, Sarah Bridle$^1$, Peter Doel$^1$, David Brooks$^1$,
\newauthor Sue Worswick$^1$, 
Gary Bernstein$^3$, 
Rebecca Bernstein$^4$, 
Darren DePoy$^5$, 
\newauthor
Brenna Flaugher$^6$, 
Joshua A. Frieman$^{6,7,8}$, 
Michael Gladders$^{7,8}$, 
Gaston Gutierrez$^6$, 
\newauthor
Bhuvnesh Jain$^3$,
Michael Jarvis$^3$, 
Stephen M. Kent$^6$, 
Ofer Lahav$^1$,
\newauthor
Aaron Roodman$^9$,
Alistair R. Walker$^{10}$\\
$^{1}$Department of Physics \& Astronomy, University College London, Gower Street, London, WC1E 6BT, UK\\
$^{2}$Institute of Cosmology and Gravitation, University of Portsmouth, Burnaby Road, Portsmouth PO1 3FX, UK\\
$^{3}$Department of Physics and Astronomy, University of Pennsylvania, Philadelphia, PA 19104, USA\\
$^4$Department of Astronomy and Astrophysics, 1156 High Street, UCO/Lick Observatory, University of California, Santa Cruz,\\ CA 95064, USA\\
$^5$Department of Physics,  Texas A\&M University, 4242 TAMU, College Station, TX 77843-4242, USA \\
$^6$Center for Particle Astrophysics, Fermi National Accelerator Laboratory, Batavia, IL 60510, USA\\
$^7$	Kavli Institute for Cosmological Physics, The University of Chicago, 5640 South Ellis Avenue, Chicago, IL 60637, USA\\
$^8$Department of Astronomy and Astrophysics, The University of Chicago, 5640 South Ellis Avenue, Chicago, IL 60637, USA\\
$^9$SLAC National Accelerator Laboratory, Menlo Park CA 94025, USA\\
$^{10}$Cerro Tololo Inter-American Observatory, National Optical Astronomy Observatory (NOAO), Casilla 603, La Serena, Chile.
}

\begin{document}

\date{}

\maketitle

\label{firstpage}

\begin{abstract}
Telescope Point Spread Function (PSF) quality is critical for realising the potential of cosmic weak lensing observations to constrain dark energy and test General Relativity. In this paper we use quantitative weak gravitational lensing measures to inform the precision of lens optical alignment, with specific reference to the Dark Energy Survey (DES). We compute optics spot diagrams and calculate the shear and flexion of the PSF as a function of position on the focal plane. For perfect optical alignment we verify the high quality of the DES optical design, finding a maximum PSF contribution to the weak lensing shear of 0.04 near the edge of the focal plane. However this can be increased by a factor of approximately three if the lenses are only just aligned within their maximum specified tolerances. We calculate the E and B-mode shear and flexion variance as a function of de-centre or tilt of each lens in turn. We find tilt accuracy to be a few times more important than de-centre, depending on the lens considered. Finally we consider the compound effect of de-centre and tilt of multiple lenses simultaneously, by sampling from a plausible range of values of each parameter.  We find that the compound effect can be around twice as detrimental as when considering any one lens alone. Furthermore, this combined effect changes the conclusions about which lens is most important to align accurately. For DES, the tilt of the first two lenses is the most important.
\end{abstract}


\section{Introduction}
Weak lensing cosmic shear has great potential to be one of the most powerful tools available to uncover the nature of dark energy \citep{detf,esoesa}. A number of planned and forthcoming surveys plan to use this probe of cosmology, including imminent surveys 
(KIlo-Degree Survey: KIDS,  
Hyper Suprime-Cam (HSC) survey\footnote{http://www.naoj.org/Projects/HSC/HSCProject.html}
and the Dark Energy Survey: DES\footnote{\tt{http://www.darkenergysurvey.org}}), telescopes under construction
(the Large Synoptic Survey Telescope: LSST\footnote{\tt{http://www.lsst.org}}), and future space telescopes
(Euclid\footnote{\tt{http://sci.esa.int/euclid}} and WFIRST\footnote{\tt{http://wfirst.gsfc.nasa.gov}}).

\begin{figure}
	\centering
	\includegraphics[width=80mm]{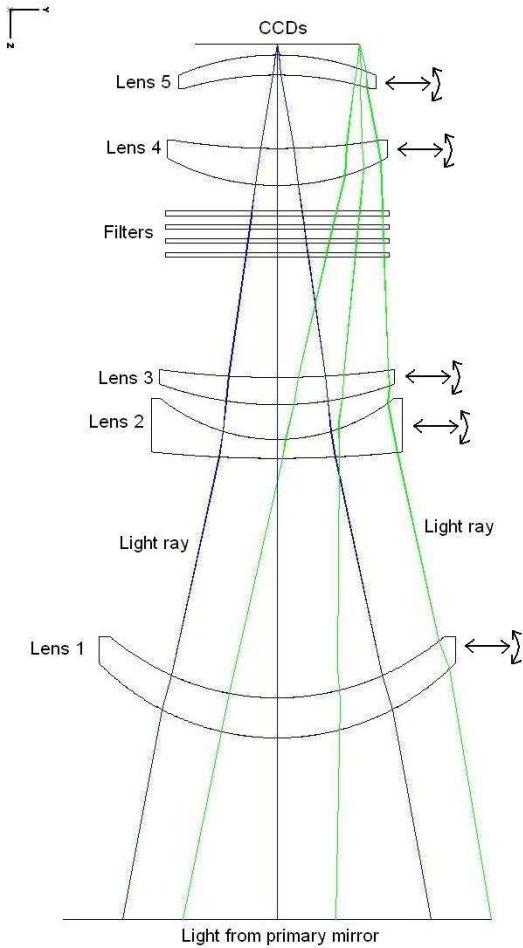}
		\caption{Diagram of the optical layout for DECam, shown in the orientation it will take when at the prime focus on La Blanco.  Two light rays are coming from the primary mirror below the optical corrector. The blue light ray is an on-axis beam and the green ray is an off-axis beam.  In our study, the lenses are decentred in the y direction and tilted around the z direction.}
		\label{fig:LensLayout}
\end{figure}

In the case of DES, weak lensing is one of four independent methods which will be used  to determine the dark energy equation of state parameter, \textit{w}, to a precision of better than 5$\%$.  The three other methods are galaxy cluster surveys, galaxy angular clustering and supernovae light curves. Details have been described in \cite{NOAO}.

In weak lensing, galaxy shapes are distorted by the curvature of intervening space-time caused by matter in the Universe.  As the majority of matter is dark and therefore is difficult to see by other means, lensing provides a useful tool.  The gravitational lensing effect can usually be described by a slight squashing of galaxy images, called shear, and a slight bending, called flexion.  The vast majority of effects are extremely small; for example an intrinsically circular galaxy would typically be sheared into an ellipse with a major to minor axis ratio of about 1.01.

Weak lensing directly probes the gravitational potential along a line of sight. Unfortunately, the atmosphere and telescope imaging also distort galaxy images, and this is usually a much bigger effect than the gravitational lensing effect we are trying to measure.   The most important effect of the atmosphere and telescope is well described by a convolution of the image with the point spread function (PSF). The PSF can be measured from images of point sources (stars); if the PSF is well known then it can in principle be removed, allowing a noisy estimate of the galaxy shear and flexion to be recovered.
However, in practice it often leaks into the lensing measurements \citep[e.g. due to model bias or noise bias][]{voigtb10,kacprzakzrbravh12}.

A key concern in developing the Dark Energy Survey programme, then, is ensuring that the Dark Energy Survey camera, DECam, has an optical system which does not add substantial image distortions.  DECam has been constructed, with careful optical alignment carried out. Installation is now taking place during the first half of 2012, and the survey is due to start in Autumn 2012. The large lens size and weight make building work challenging due to the tight tolerances on the positioning.  These tolerances have been designed to keep the instrumentation distortions to a minimum, so that their contributions to the weak lensing systematics is minimal, but even so, the expected ellipticity of the PSF could be larger than the lensing signal, so careful modelling of the optical distortions will be needed to recover cosmological information.

In order to ensure that aspects of the alignment have been prioritised in proportion to the influence they will have on weak lensing systematics, we have engaged in the studies presented in this paper. We  examine what impact errors in lens alignments have on the camera contribution to the lensing signal. The work shown here is specific to the DECam optical design, but also acts as a case study for development of future lensing-optimised optical telescopes. This work follows on from earlier work on PSF requirements from lensing by \citet{2006SPIE.6269E.100K}, who computed PSF distortions for DES for a variety of different contributions from the optics and compared the result with those from other telescopes. 

This paper is organised as follows. In section~\ref{sec:optics} we review the design of DECam. We discuss the shear and flexion measures used to evaluate the optical performance in section~\ref{sec:metrics}. We present our results in section~\ref{sec:results}, and discuss the implications for lensing measurements in section~\ref{sec:final}.

\begin{figure*}
			\centering
			\includegraphics[width=14cm]{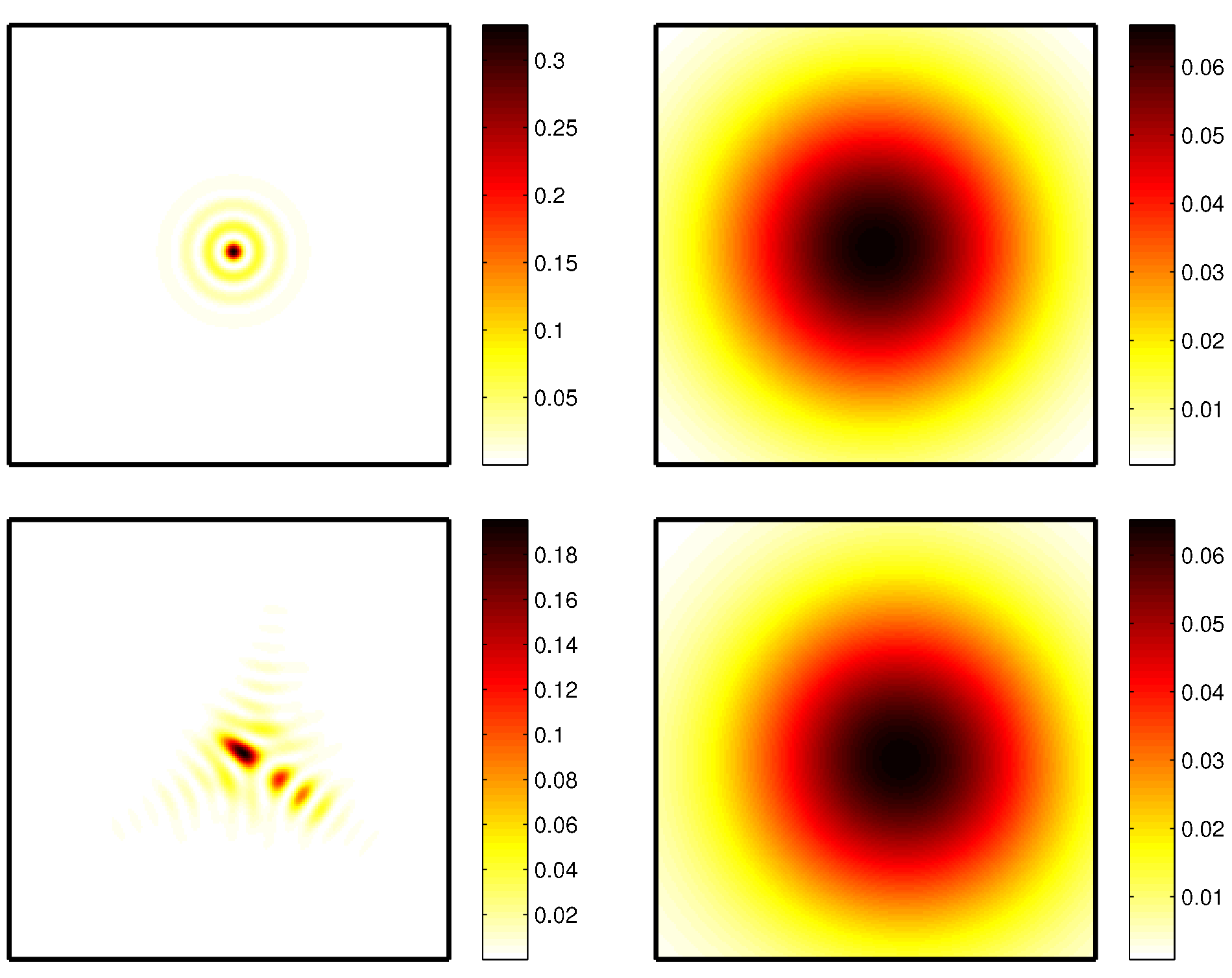}
		\caption{The camera point spread function at the centre of the focal plane (top left), and at the edge of the focal plane (bottom left).  Right panels show the corresponding PSFs combined with a $0.7''$ FWHM Gaussian, to describe the PSF including atmospheric seeing.}
		\label{fig:PSFs}
\end{figure*}

\begin{table}
	\centering
\begin{tabular}{|c|c|c|c|c|}
	\hline
	 & 
	\multicolumn{2}{|c|}{Assembly Tolerances} & 
	\multicolumn{2}{|c|}{Dynamic Tolerances}   \\
	  &  De-centre  &  Tilt Tolerance  &  De-centre  &  Tilt Tolerance  \\
	  &  Tolerance  &  on diameter  &  Tolerance  &  on diameter  \\
	 Lens &  ($\mu$m)  &  arcsec ($\mu$m)  &  ($\mu$m)  &  arcsec ($\mu$m)  \\
	\hline  
	  C1  &  100 &  10 (48)  &  25  &  5.6 (27)  \\
	  C2  &  50 &  17 (56)  &  25  &  8.1 (27)  \\
	  C3  &  100 &  20 (63)  &  25  &  8.4 (27)  \\
	  C4  &  100 &  20 (58)  &  25  & 8.6 (25)  \\
	  C5  &  200 &  40 (105)  &  25  &  10 (25)  \\
	\hline
\end{tabular}
	\caption{De-centre and alignment tolerances of the lenses for assembly and stability during operation, from Doel et al. 2008.
	} 
\label{tab:tolerances}	
\end{table}

\begin{table}
	\centering
		\begin{tabular}{|c|c|c|c|c|}
			\hline
		  & Centre Thickness  & Edge Thickness & Diameter & Weight \\
		Lens & (mm) & (mm) & (mm) & (kg) \\
				\hline
		C1 & 112.1 & 74.34 & 980 &  172.36  \\
		C2 & 51.285  & 148.11 & 690 & 87.20   \\
		C3 & 75.1  & 38.27 & 652 & 42.62  \\
		C4 & 101.68 & 52.48  & 604 & 49.69  \\
		C5 & 54.68  & 36.19  & 542 & 24.37 \\
			\hline
	\end{tabular}
\caption{Dimensions of the DECam lenses, from  Doel et al. 2008.}
\label{tab:lensSize}
\end{table}

\section{Optical Design of DECam}
\label{sec:optics}

DECam is a new wide-field prime focus camera. The optics consist of five lenses, ranging in  diameter from 980mm to 542mm~\citep{bib:Antonik2009,bib:Brenna2010_mnras}, with six filters between the third and fourth lenses covering wavelengths of 350 to 1100nm, the layout of which is shown in figure \ref{fig:LensLayout}.  Lens 5 acts as the window to the cryostat that holds the CCDs; this cooling reduces CCD noise.  Lenses 1 to 4 are held in nickel-iron cells~\citep{bib:PeterMarseille} glued into position using a silicone rubber solution.  Lens 5 is held in its cell during operation by the vacuum under which it is placed.  These cells are attached to the barrel which is supported in the primary cage above the primary mirror.

The construction of the optical corrector includes two main alignment stages: (i) the alignment of the lenses into their cells and (ii) the alignment of the lens-cells into their holding barrel, which is held at prime focus.  In order for precisely focused images to be created, each part of the optical corrector must be aligned to within a series of carefully defined tolerances including (i) manufacturing tolerance; (ii) ``assembly tolerance'' to which the element must be aligned with respect to the nominal optical axis of the camera, when the optical corrector is being built; and (iii) the acceptable ``dynamic tolerance'' amount by which the lenses can move as the optical corrector is moved across the sky.  The optical corrector is designed to minimise movement once assembled; however, such motions cannot be entirely suppressed and so during operation the optical alignment will change slightly depending on the position of the telescope relative to the zenith.  Tolerances to which the lenses must be aligned are given in Table \ref{tab:tolerances}.  These tolerance are challenging to meet due to the size and weight of the lenses, dimensions of which are given in Table \ref{tab:lensSize}.  

This paper focuses on the effect of non-perfect alignment of the lenses which occurs during the lens-cell assembly stage, when the lenses are glued into their holding cells.  While lenses were not glued until they are within tolerance, perfect alignment is not possible within any realistic time frame.  It has therefore been useful to discover which lenses affect the weak lensing data the most, to ensure higher precision of alignment was attempted for these lenses.  In the following sections we explore which lenses need the most attention to increase the science output of the camera.

\begin{figure*}
			\centering
			\includegraphics[width=14cm]{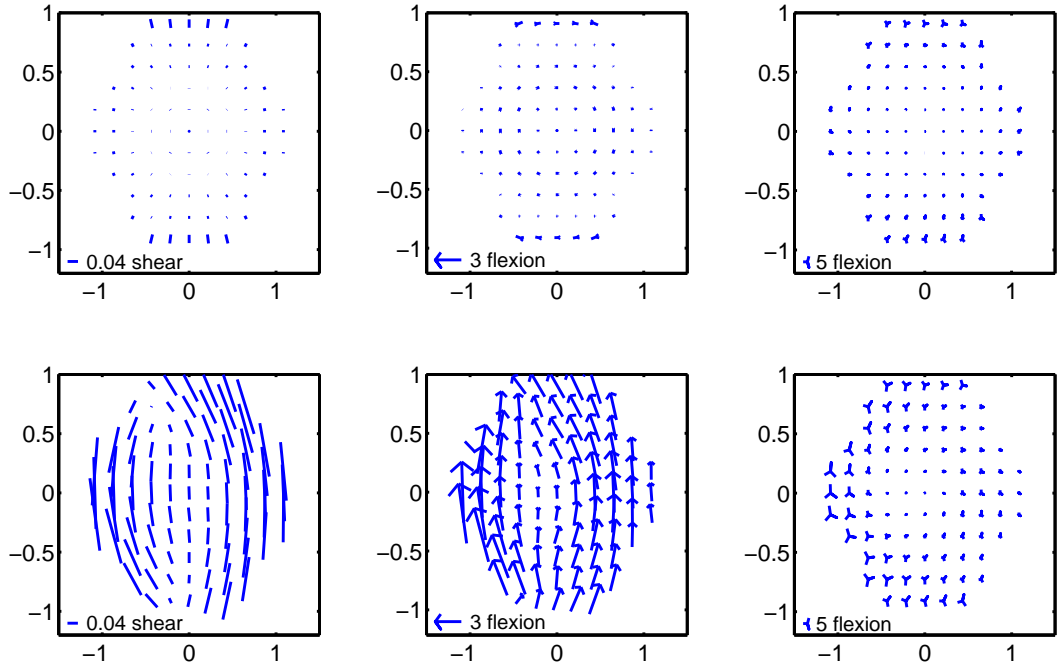}
		\caption{Whisker plots for the shear, 1- and 3- flexion, distributed over the full focal plane.  Bottom left corner shows the scale of the distortion; flexion is in units of arcsec$^{-1}$.
		}
		\label{fig:BestWorstWhiskers}
\end{figure*}

\section{Shear and Flexion}
\label{sec:metrics}

In this section we describe the requirements and goals we set for the telescope and atmospheric point spread functions (PSF).  We assume throughout that the atmosphere and CCDs contribute a circular Gaussian component of Full-Width Half Maximum (FWHM) 0.7 arcseconds, representing good seeing conditions at Cerro Tololo where the median with the Blanco telescope together with the Mosaic imager is 0.9 arcseconds \citep{NOAO}. We convolve this with the optics point spread function to obtain the full PSF which we consider.

To perform a weak lensing analysis, a galaxy image is corrected to remove the impact of the PSF as well as possible. However, several effects can cause an imperfect PSF removal, for example insufficient knowledge of the PSF~\citep[e.g.][]{paulinavrb08} and a wrongly assumed model for the galaxy~\citep{voigtb10}. These PSF residuals will propagate into the galaxy shear estimates and contaminate the cosmological analysis~\citep[e.g.][]{Amara:2007as}. The amount by which the residuals propagate depends on the galaxy shear measurement method used, as demonstrated in a series of weak lensing data analysis challenges~\citep{STEP1mnras,STEP2mnras,great08resultsmnras}. Therefore in this paper we mainly consider the raw PSF ellipticity shear and flexion measures, and only consider later the probable fraction of this which might leak into weak lensing estimates.

We generate PSF convolution contributions for the atmosphere and the instrument separately.  A simple circular Gaussian is used for the atmosphere. The instrument PSF depends on the optical layout, which we input into the optical modelling program ZEMAX\footnote{\tt{http://www.zemax.com}}.
This program allows simulated light rays to be traced through an optical system and produces a resulting PSF at a given position in the focal plane.  For all the following results the central ray of the i band filter, at a wavelength of 743.9nm, was used.  We defer discussion of the impact of the wavelength dependence of the DES PSF~\citep{cyprianoavbars10,2012MNRAS.421.1385V} to future work. At all times the focal plane remained in the same position at perfect alignment.

\begin{figure*}
			\centering
			\includegraphics[width=14cm]{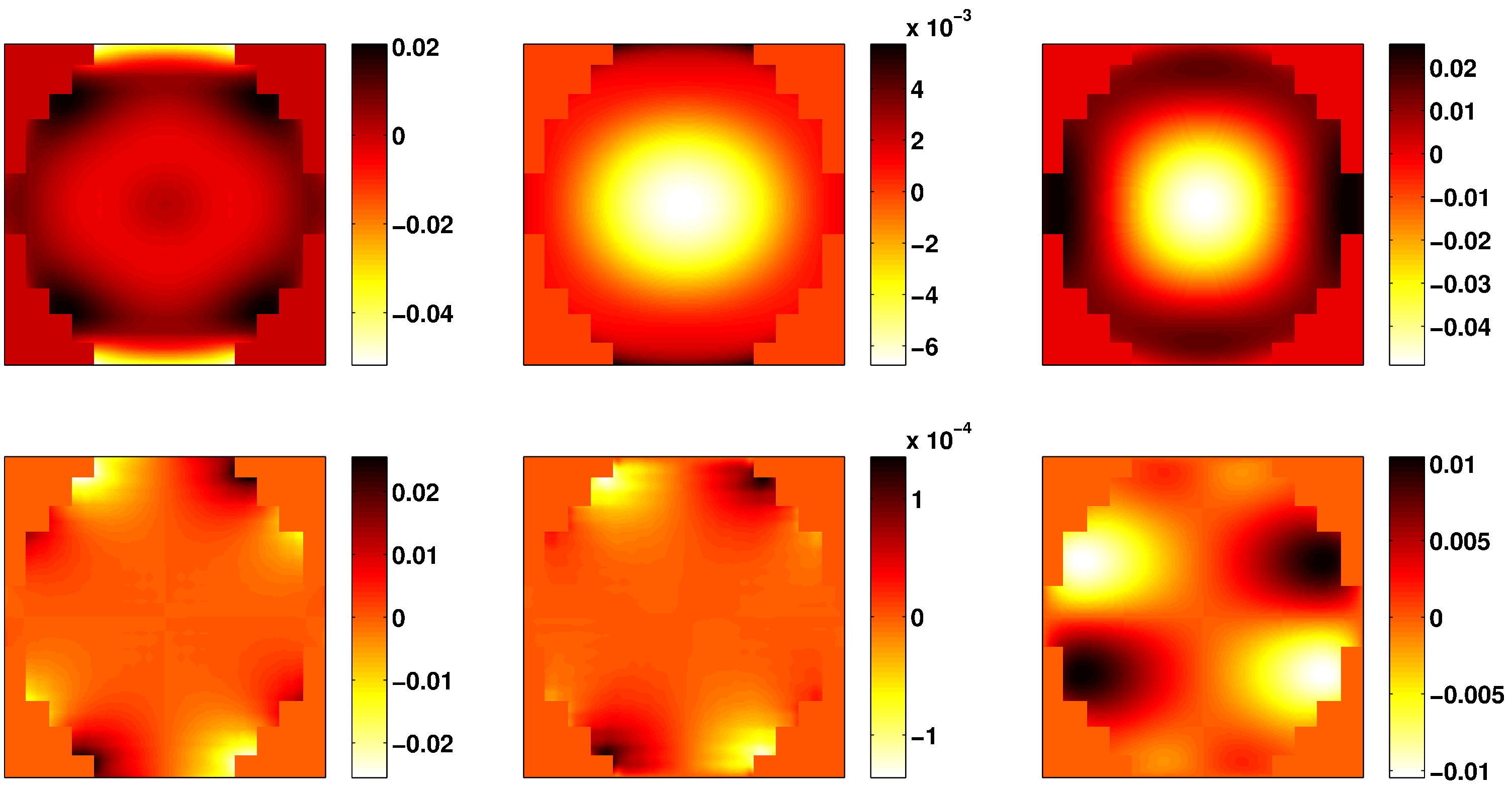}
		\caption{Convergence maps for the shear (left panels), 1-flexion (middle panels) and 3-flexion (right panels) for a perfectly aligned optical system. Upper panels show the E modes and lower panels show B modes. Flexion is in units of arcsec$^{-1}$.}
		\label{fig:KappaMaps}
\end{figure*}

Example images of the generated PSFs are shown in Figure \ref{fig:PSFs}, before and after they are combined with a Gaussian atmosphere, at positions 0,0 and -0.54,0.46 degrees from the centre of the field.  Moments were then used to convert the PSFs into shear and flexion measures using 
first moments
\bea
\bar{x} &=& \frac{\int \, I(x,y) \, x \, \rmd x \, \rmd y}
{\int \, I(x,y) \, \rmd x \, \rmd y}\\
\bar{y} &=& \frac{\int \, I(x,y) \, y \, \rmd x \, \rmd y}
{\int \, I(x,y) \, \rmd x \, \rmd y} ~,
\eea
quadrupole moments
\bea
\mathcal{Q}_{xx} &=& \frac{\int I(x,y) \, (x-\bar{x})^2 \, \rmd x \, \rmd y}
{\int \, I(x,y) \, \rmd x \, \rmd y} \label{eqn:qxx}\\
\mathcal{Q}_{xy} &=& \frac{\int I(x,y) \, (x-\bar{x})(y-\bar{y}) \, \rmd x \, \rmd y}
{\int \, I(x,y) \, \rmd x \, \rmd y}\label{eqn:qxy}\\
\mathcal{Q}_{yy} &=& \frac{\int I(x,y) \, (y-\bar{y})^2 \, \rmd x \, \rmd y}
{\int \, I(x,y) \, \rmd x \, \rmd y}.\label{eqn:qyy}
\eea
and the further moments given by \cite{okura} required for 1- and 3-flexion estimates. The ellipticity can then be found for a noise-free image using
\be
\epsilon \equiv \epsilon_1+i\epsilon_2= \frac{\mathcal{Q}_{xx} - \mathcal{Q}_{yy}+2i\mathcal{Q}_{xy}}
{\mathcal{Q}_{xx}+\mathcal{Q}_{yy}+2(\mathcal{Q}_{xx}\mathcal{Q}_{yy} - \mathcal{Q}_{xy}^2)^{1/2}} ~,
\ee
\citep{bonnetm95}
where we introduce the standard complex number notation $\epsilon = \epsilon_1 + i \epsilon_2$ where $i^2=-1$. 
An intrinsically circular galaxy becomes stretched into an ellipse with ellipticity $\epsilon$ on application of a shear $\gamma$. We will apply the same notation to PSFs and use the above equation to calculate the `shear' of the PSF. We discuss how this impacts on galaxy shear measurements in section 4.

The corresponding equations relating moments to flexions are provided by \cite{okura}, equations (23)-(24). Similarly to shear, we use these equations to calculate the flexion of the PSF.

\begin{table*}
	\centering
		\begin{tabular}{|c|c|c|c|c|c|c|}
			\hline
			  & $\sigma_{E}$ shear &  $\sigma_{B}$ shear & $\sigma_{E}$ 1-flexion & $\sigma_{B}$ 1-flexion & $\sigma_{E}$ 3-flexion &  $\sigma_{B}$ 3-flexion \\
			  \hline
			 Perfect case & 0.0086 & 0.0044 & 0.0026 & 2.29E-5 & 0.0168 & 0.0038 \\
			 Worst case & 0.0557 & 0.0244 & 0.0093 & 0.0049 & 0.0317 & 0.0276 \\
			 \hline
	\end{tabular}
\caption{Standard deviation of the convergence maps for shear and flexion for perfect alignment and worst case alignment. Flexion is in units of arcsec$^{-1}$.}
\label{tab:SigmasCompare}
\end{table*}

We show example shear and flexion whisker plots
in Fig.~\ref{fig:BestWorstWhiskers}.
The upper panels use the best-case alignments, and we can see that the shear is everywhere less than 4 per cent and the flexion is
less than 3 per arcsec. 
For comparison, we also show the results for a ``worst-case'' misalignment, as defined during the optical design: that giving the largest RMS spot size. 
This has a large area with shear values greater than 0.1, reaching 0.2 at the edges, with a significant tangential component which would indicate a spurious mass concentration in the center of the field if unaccounted for. The flexion maps have values up to 5 per arcsec,
and generally shows a displacement in a single direction for the 1-flexion, and a trefoil angled in a similar pattern to the perfect alignment case for the 3-flexion.

These plots of PSF distortion as a function of position on the focal plane
can be used to 
reconstruct the projected matter density map that would be obtained if they were uncorrected, using extensions of the \cite{ks} reconstruction method.
This gives both real and imaginary density distributions, depending on the orientations of the distortions. This is because lensing by matter usually only produces particular patterns of distortions. It is frequently assumed that the imaginary density map from real data gives an indication of the amount of systematic e.g. telescope effects which have leaked into the real density map. It is therefore of considerable interest to examine whether typical telescope distortions produce similar amounts of real and imaginary contributions to the density map, and so we consider both real and imaginary density maps here.
For shear the required equations are the usual \cite{ks} results 
\begin{equation}
\tilde{\kappa}_{S}(\vec{\textit{k}}) = \tilde{\gamma}(\vec{\textit{k}})\tilde{D}(\vec{\textit{k}})
\end{equation}
where $\tilde{\kappa}_{S}$ is the Fourier space scaled projected density (convergence) map, 
$\tilde{\gamma}$ is the Fourier transform of the PSF shear map and
$\tilde{D}$ 
is the response of a delta function shear map
\begin{equation}
\tilde{D} = \frac{k_{1}^{2} - k_{2}^{2} - 2ik_{1}k_{2}}{k_{1}^{2} + k_{2}^{2}} .
\end{equation}
We extend the Kaiser \& Squires formalism to flexion and find 
\begin{equation}
\tilde{\kappa_{F}}(\vec{\textit{k}}) = \tilde{F}(\vec{\textit{k}})\tilde{Q}(\vec{\textit{k}})
\end{equation}
\begin{equation}
	\tilde{Q} = \frac{-ik_{1} - k_{2}}{k_{1}^{2} + k_{2}^{2}}
\end{equation}
and
\begin{equation}
\tilde{\kappa_{G}}(\vec{\textit{k}}) = \tilde{G}(\vec{\textit{k}})\tilde{R}(\vec{\textit{k}})
\end{equation}
\begin{equation}
	\tilde{R} = \frac{-i(k_{1}^{3} - 3k_{1}k_{2}^{2}) + (k_{2}^{3} - 3k_{1}^{2}k_{2})}{(k_{1}^{2} + k_{2}^{2})^2} .
\end{equation}
Fig.~\ref{fig:KappaMaps} shows the resulting convergence reconstructions for the best case PSF shear and flexion maps. The PSF shear is roughly circularly symmetric, with tangential distortions in the centre surrounded by radial distortions. It therefore makes sense that the E-mode density reconstruction (top left panel of Fig.~\ref{fig:KappaMaps}) is positive in the centre with a surrounding negative ring. Since shear is non-locally related to density, the central tangential shears are also partially accounted for by positive density around the edges of the field, which occurs especially at the diagonal corners of the map. The B-mode density reconstruction from the shear map is an indicator of the extent to  which the shear map is not circularly symmetric. In the centre part of the field the range in the B-mode density is slightly smaller than that for the E-mode density, as can be expected by the symmetry of the perfect alignment case. The non-trivial relationship between the size of the E-mode density and B-mode density across the field is particularly interesting in the context of systematics assessments in cosmic shear, which often assume a similar amplitude for both modes. 

The 1-flexion translates into a density map which is an order of magnitude smaller than that from the shear, which bodes well for use of 1-flexion for analysing true density peaks.  The density range for the 3-flexion is similar to that for shear, with a significant negative peak in the center of the field. This is expected from the radially growing trefoil shape of the PSF seen in the top right panel of Fig. 3.

\begin{figure*}
			\centering
	\includegraphics[width=14cm]{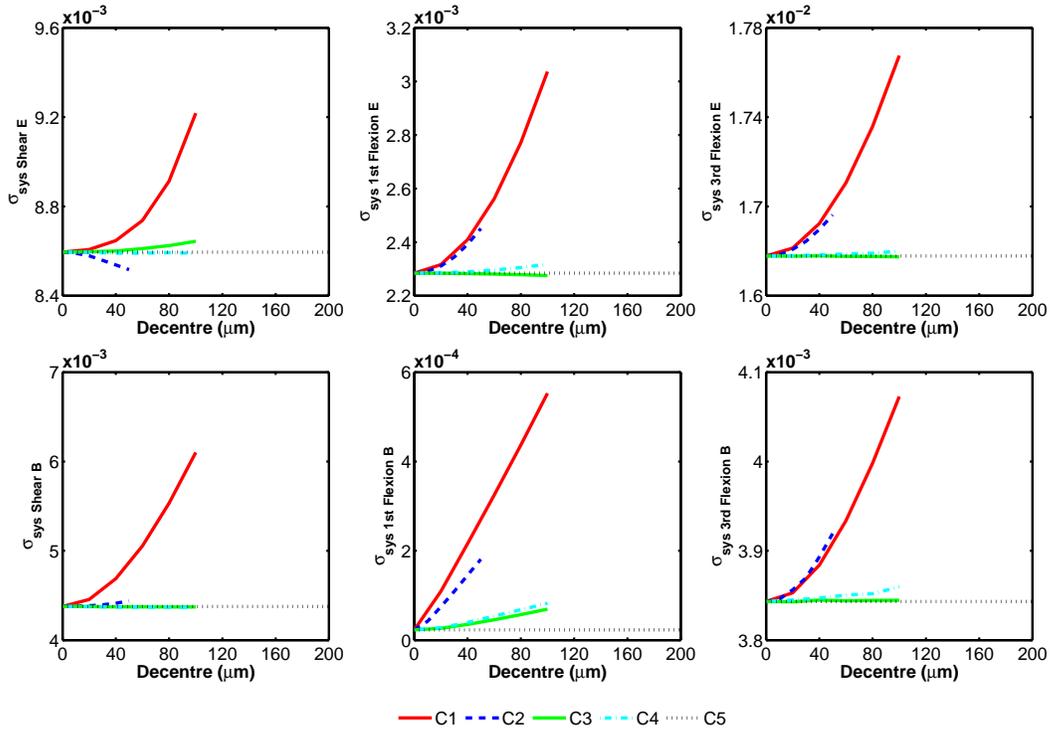}
		\caption{The rms real part (top) and imaginary  part (bottom) of the projected density reconstructed from the PSF shape component, as a function of de-centre of each lens individually.  The left hand panels are for shear, the center panels are 1-flexion and the right hand panels are 3-flexion.  Flexion is in units of arcsec$^{-1}$.
		 }
		\label{fig:SigmaEDecentre}
\end{figure*}

\begin{figure*}
			\centering
			\includegraphics[width=14cm]{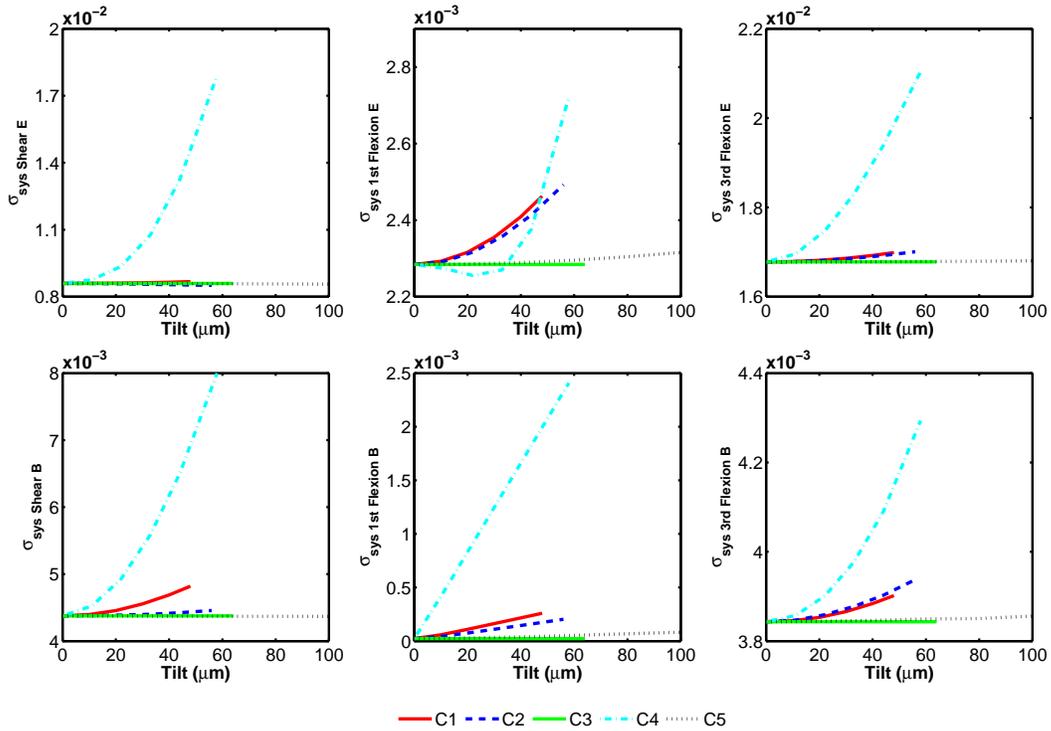}
		\caption{As Fig. 5, but as a function of the tilt of each individual lens.}
		\label{fig:SigmaETilt}
\end{figure*}

\begin{figure*}
	\centering
		\includegraphics[width=14cm]{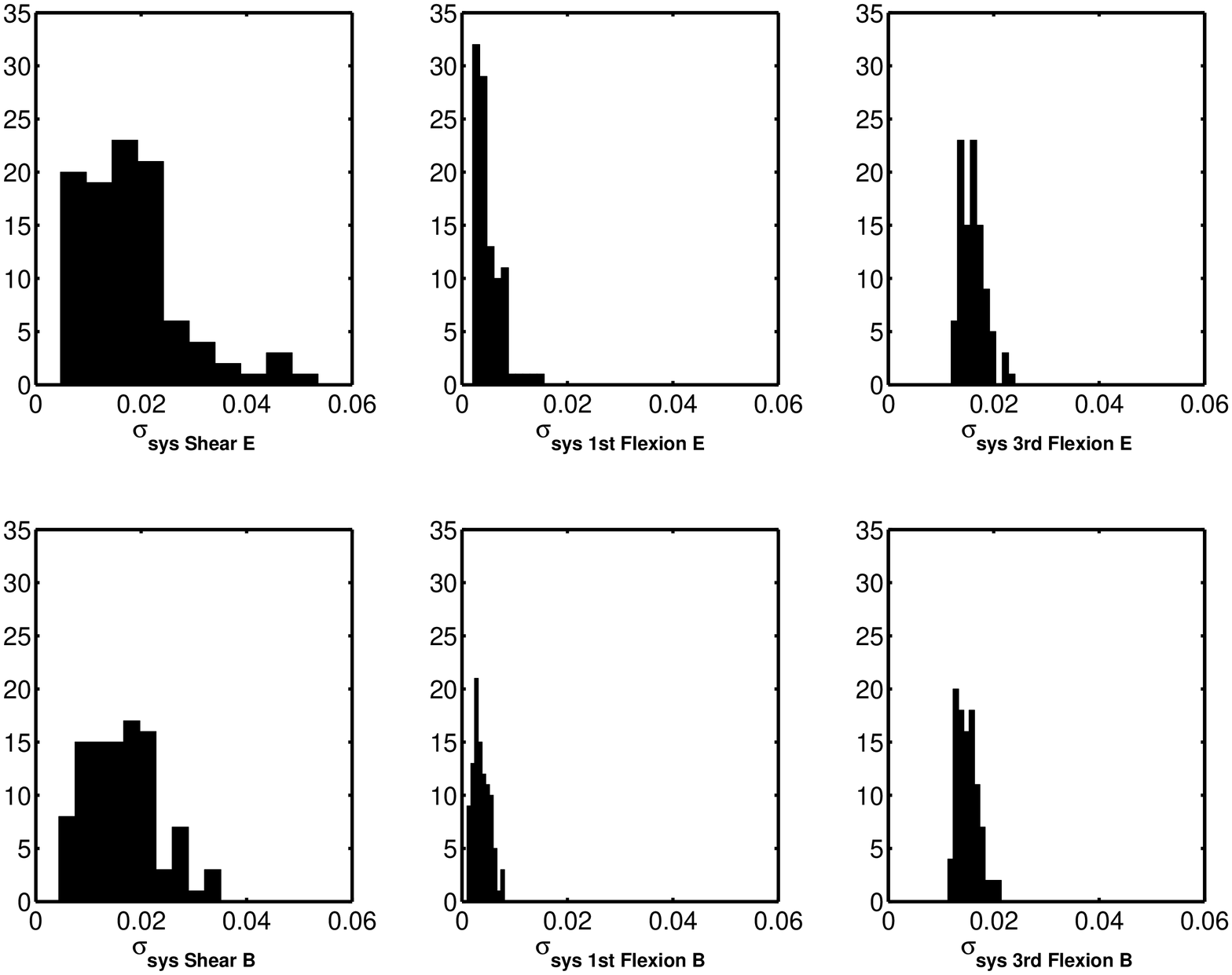}
		\caption{Histogram of 63 random misalignments of all lenses, showing $\sigma_E$ and $\sigma_B$ of the system.  Left panel is shear, centre panel is 1-flexion, and right panel is 3-flexion.  Flexion is in units of arcsec$^{-1}$.
		 }
		\label{fig:MockHistogram}
\end{figure*}

In order to encapsulate the convergence maps into a single number for comparisons between camera configurations, we use the standard deviation, $\sigma_{\rm sys}$, of the convergence map. This statistic is important for statistical analyses of shear and flexion, for example cosmic shear correlation functions. Standard deviations are given in Table~\ref{tab:SigmasCompare} for the perfect alignment and worst case alignments.  As expected from the density map (Figure 4), the largest standard deviations for perfect alignment are for the E-mode shear and E-mode 3-flexion. 
For both perfect alignment and worst case alignment, the standard deviation for E-mode shear and B-mode shear are a comparable size, which suggests that systematics checks of the B-mode are a useful diagnostic of PSF leakage in the E-mode signal. However, the B-modes are a factor of two smaller than the E-modes, even for worst case alignment, so the exact value of the B-mode signal cannot be taken as an estimate of E-mode contamination (e.g. cannot be simply subtracted off the E-mode to find a true E-mode). The values for 1-flexion are still very small even for the worst case alignment, while the 3-flexion becomes large for both E- and B- mode in the worst case.

\section{Effects of Misalignment}
\label{sec:results}

In this section we examine the effect of lens misalignments on the image quality using two methods. We first move each lens in turn with the others remaining in perfect alignment.  This allows us to observe the effect which each lens has on the system.  Secondly, we move all the lenses by a random amount, whilst staying within the given tolerances, to see the effect of multiple misalignments.  Looking at multiple lens displacements allows an examination of typical shear values which the system will exhibit.  

\subsection{Relative effect of single lenses}

We first de-centre each lens in turn while keeping all other lenses in perfect alignment. 
The resultant shears and flexions of the PSF generated are then converted into E and B mode convergence maps, and we calculate standard deviations $\sigma_{E}$ and $\sigma_{B}$.  
These are plotted as a function of de-centre distance in Fig.~\ref{fig:SigmaEDecentre}. 
The $x$-axis is the de-centre distance in microns, and the lines terminate at the tolerance criteria.
Note that the $y$-axis does not start at zero, and thus we note that for the quantities generating the largest convergence variances (shear and 3-flexion) the exact misalignment value of any one lens does not change the convergence variance by a very large amount.

We see that lens C1 has the biggest effect of any single lens which is not suprising as it is highly curved and at the start of the optical system.
Lens C2 also has a comparable effect on 1- and 3-flexion, while the other lenses have an effect on the variance more than ten times smaller than C1.

\begin{figure*}
			\centering
			\includegraphics[width=14cm]{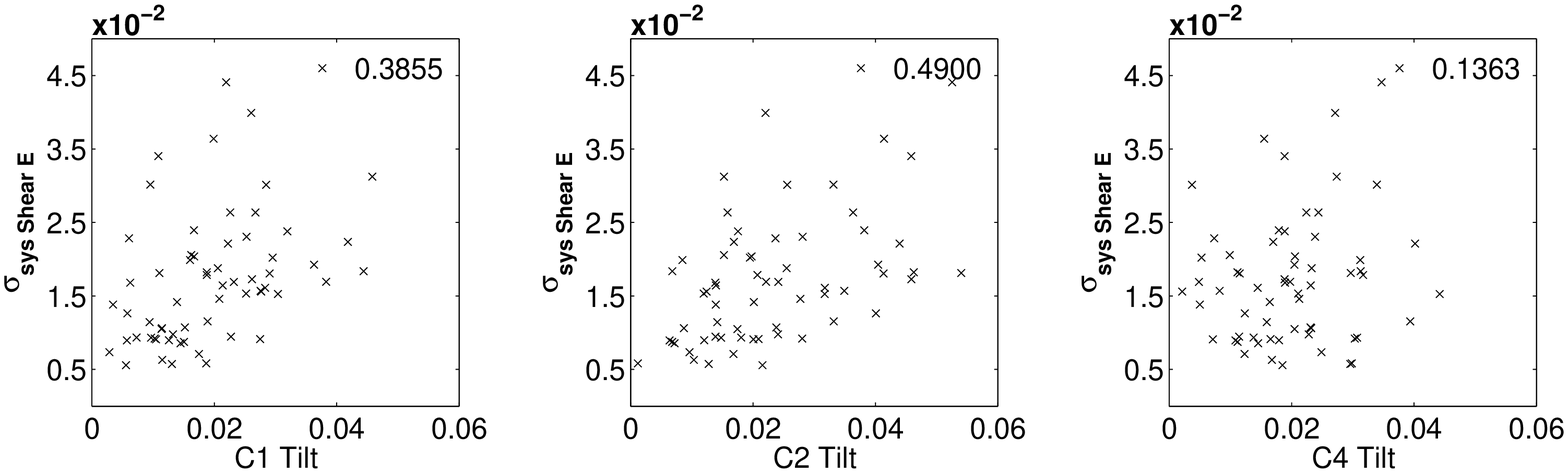}
		\caption{Plots of the systematic $\sigma_E$ shear against the tilts on C1, C2 and C4.  The value in the top right hand corner is the correlation coefficient. Tilt is in units of mm.}
		\label{fig:SigmaSysTiltCorr}
\end{figure*}

We next consider a tilt of the lenses in Fig. \ref{fig:SigmaETilt}. We see that the lens tilts are much more important for lensing measurements than the lens de-centres, especially the tilt of the lenses with aspheric surfaces C2 and C4. For the most important cosmological quantity, E-mode shear, the convergence variance roughly doubles as the C4 tilt goes from perfect alignment to the full tolerance range, while tilt of other individual lenses  has relatively little impact.
The E-mode 1-flexion is not greatly changed from its original small value by the tilts, and although the B-mode first flexion increases by a large factor especially due to C4 tilt, the value is still small. C4 also has the biggest effect on the third flexion, and C1 and C2 have considerably smaller effects. Both C3 and C5 have little effect when tilted or decentred.

\subsection{Multiple lens offsets}

We now wish to answer two questions: (i) Does the relative importance of the lenses hold when more than one lens is offset from the ideal position? (ii) What is a realistic size of distortion that will come from the completed optical camera?  In order to answer these questions we sampled 100  different alignments from a multi-variate Gaussian distribution of de-centres and tilts, 
where the standard deviation for each de-centre or tilt is equal to one third of the tolerance. 63 of these landed within the tolerance boundaries for the camera alignment, and we will consider this sample.

The resulting convergence standard deviations are plotted as a histogram in Fig.~\ref{fig:MockHistogram}.  The most important quantity for cosmology, the shear E-mode convergence range, is shown in the top-left panel. It has the widest range of all the quantities, which extends by more than a factor of two beyond the values obtained by varying the alignment of any one lens. It also extends below the perfect alignment case, presumably due to certain cancellations. A similar but slightly less extreme pattern exists for the B-modes from the PSF shears. The flexion histograms are quite tight around the perfect alignment case, and thus place less pressure on the need for stringent alignments.

Next we investigate which lens is causing the greatest distortions, in this multi-parameter variation. We plot the convergence standard deviation $\sigma_{\rm sys}$  against the tilt and de-centre of each lens. The most significant examples are shown in Fig.~\ref{fig:SigmaSysTiltCorr} for the misalignments which had the biggest effect when varied individually. 
We calculate the correlation coefficient between the $\sigma_{\rm sys}$ values and the tilts and decenters of each lens, showing the results in Table \ref{tab:RandomVaules}.  By comparing with random samples from Gaussian uncorrelated distributions with the same number of samples, we find that a value above 0.26
means that the lens gives a significant contribution to either the shear or flexion, at 95\% confidence. 
It is interesting to note that  C4 tilt, which looked very important in figure \ref{fig:SigmaETilt}, has  little correlation with the $\sigma_{\rm sys}$ of the multiple lens off-sets.  However, the tilt of lenses C1 and C2 now dominate the distortions. 
This is in keeping with the previous DECam design observation
that C1 and C2, when displaced, have the largest impact on the size of the root mean square spot size of the PSF.

\begin{table}
\begin{tabular}{|c|c|c|} 	\hline
			&   correlation co-effiecent &   correlation co-effiecent \\
	    &   E mode & B mode  \\   \hline
	    C1 decentre &  &   \\
	    Shear  &  -0.1301 &  -0.1123 \\
	  First Flexion  &  -0.1621&  -0.2183   \\
	  Third Flexion  &  -0.0680 &	-0.0254  \\	 \hline
	  C2 decentre &  &  \\
	  	    Shear  & -0.0715  & -0.0575 \\
	 First Flexion  & -0.1032 &  -0.1289 \\
	  Third Flexion  & -0.0731 &	-0.1084 \\	  \hline
	  C3 decentre &  &  \\
	  	   Shear  & -0.0631 &  0.0016 \\
	  First Flexion  & -0.0692 &  0.0025 \\
	  Third Flexion  & -0.0291 &	 0.0127 \\	  \hline
	  C4 decentre &  &  \\
	  	    Shear  & 0.0635  &  0.0724 \\
	  First Flexion  &    0.0740  &  0.0402 \\
	 Third Flexion  &  0.1141 &	0.0750  \\	  \hline
	 C5 decentre &  &  \\
	  	    Shear  & -0.4317 &  -0.3798 \\
	  First Flexion  & -0.2357  &  -0.2920 \\
	  Third Flexion  & -0.4157  &	-0.3952 \\	  \hline
	  C1 tilt &  & \\
	  	   Shear  &  0.3855 & 0.5490 \\
	  First Flexion  &  0.5490 & 0.6884 \\
	   Third Flexion  &  0.6884 & 0.6603 \\	  \hline
	   C2 tilt  &  &  \\
	  	    Shear  &  0.4900 & 0.6670 \\
	 First Flexion  & 0.0605 &  0.1887 \\
	 Third Flexion  &  0.2504 &	0.3660 \\	  \hline
	  C3 tilt  &  &   \\
	  	   Shear  &   0.1744 & 0.1118 \\
	  First Flexion  & 0.0551  & -0.0550 \\
	  Third Flexion  &  -0.0118 &	-0.0111 \\	  \hline
	   C4 tilt &  &   \\
	  	   Shear  & 0.1363  &   0.2107 \\
	  First Flexion  & 0.1406 & 0.0792 \\
	  Third Flexion  &  0.0459 &	0.0458 \\	  \hline
	  C5 tilt &  &   \\
	  	    Shear  & 0.2526  &  0.2218 \\
	  First Flexion  &  0.1595 &  0.1838 \\
	  Third Flexion  &    0.2150 &	0.2498 \\	  \hline
\end{tabular}
	 \caption{Table showing correlation coefficients for E-mode and B-mode distortions, from the covariance matrix for random misalignments, under the given tolerances. Flexion is in units of arcsec$^{-1}$.}
   \label{tab:RandomVaules}
\end{table}

\section{Impact on Galaxy Shear Measurement} 
\label{sec:final}

In order to recover a weak lensing signal, the systematics in the residual shear must be controlled to allow the results to be statistically significant.  This tolerance on the variance, $\sigma_{\rm sys}^{2}$ can be defined by area of the survey A$_{s}$, the galaxy density $n_{g}$, and the median redshift $z_{m}$ \citep{Amara:2007as} where
\begin{equation}
\sigma_{\rm sys}^{2} < 10^{-7}\left(\frac{A_{s}}{2\times10^{4}{\rm deg}^{2}}\right)^{-0.5}\left(\frac{n_{g}}{35 {\rm arcmin}^{-2}}\right)^{-0.5}\left(\frac{z_{m}}{0.9}\right)^{-0.6}
\end{equation}
For DES, we have A$_{s}=5000$ deg$^{2}$, n$_{g} \simeq 10$ arcmin$^{-2}$ and z$_{m} \simeq 0.7$ \citep{NOAO}.  This puts a requirement on DES for $\sigma_{\rm sys}$ to be less than 0.0008.  While all the results quoted previously are significantly above this level, this has not yet taken into account the expected correction by the weak lensing pipeline.  

From the comparison of current methods by \cite{great08resultsmnras}, the residual shear after correction would be a factor of at least 120 times smaller than the PSF shear described in this paper. Reductions in the flexions were not calculated in  \cite{great08resultsmnras}, so it is not yet known whether they can be reduced by a similar amount.  However, for shear, this factor brings $\sigma_{\rm sys shear}$ of the residual shears under the required 8x10$^{-4}$ for both perfect and worst case alignment, as shown in table \ref{tab:ResidShears}.

\begin{table}
	\centering
	\begin{tabular}{ccc}
	\hline
	   &   \multicolumn{2}{c}{Residual Shear}  \\
	   &  $\sigma{\rm sys_{E}}$ &  $\sigma{\rm sys_{B}}$ \\ 
	\hline
	Perfect alignment  & 2.7x10$^{-4}$  &  1.4x10$^{-4}$  \\
	Worst alignment   &  4.3x10$^{-4}$  &   2.5x10$^{-4}$  \\
	\hline
	\end{tabular}
	\caption[Remaining shear in the image after analysis]{Expected size of remaining shear in the image after the telescopic distortions have been removed to the extent that current analysis techniques allow.}
	\label{tab:ResidShears}
\end{table}

\section{Conclusions}

In this paper we have explored the effect of camera optical alignments on quantities important for weak gravitational lensing, with specific reference to the Dark Energy Survey (DES). 
We have calculated the shear and flexion of the PSF as a function of position on the focal plane, for a realistic raytracing model of DECam including misalignments of lenses. In this work, we do not include flexure errors, or spacing tolerances which can be overcome by adjusting focus. 
We also do not consider flatness, position or tilt of the focal plane.
However, we include lens decentres and tilts, which are known to be within tolerance for the actual DECam.

For perfect optical alignment, we find a maximum PSF shear of 0.04 near the edge of the focal plane. However this can be increased by a factor of roughly three if the lenses are only just aligned within their maximum specified tolerances. In both cases, however, once PSF correction methods are employed, the resulting impact on shear estimates for galaxies in the DES weak lensing catalogue 
are expected to be below the required threshold for measuring dark energy parameters. 

We have calculated the E and B-mode shear and flexion variance as a function of de-centre or tilt of each lens in turn. We found tilt accuracy to be a few times more important than de-centre, depending on the lens considered. Finally, we have considered the combined effect of de-centre and tilt of multiple lenses simultaneously, by sampling from the permitted range of values of each parameter.  In this case we find that combined effect can be around twice as detrimental as when considering any one lens alone. Furthermore, this combined effect changes the conclusions about which lens is most important to align accurately; for DES, the tilt of the first two lenses is the most important.

The results of these simulations have been used to inform
alignment of the DECam lenses (i.e. on which lens
alignments to spend the most effort), and acts as an important
confirmation that the DECam optical tolerances lead
to a system which is fit for extremely accurate weak lensing
measurements.
The final image quality of course will be
affected by other factors such as barrel sag, optical wedges, surface figure errors, focal plane misalignment and material inhomogeneity (in roughly decreasing order of significance). 

For example, the quality of the lens polishing introduces surface figure errors.
We have assumed perfect polishing
for this work. The smoothness of the final DES lenses have
been examined using interferometry and photography to assess
phase deviations and is well within specifications. Preliminary
investigations suggest that typical atmosphere convolved
PSF ellipticities are increased by up to twenty per
cent in a way which varies in a non-symmetric way across
the field of view. It appears to be a small but non-negligible
fraction of the effect due to random misalignments of lenses
within their tolerances studied in this work.

Work is
ongoing to incorporate all the image quality effects into the optical model
along with the actual measured mis-alignments of the optics
to produce a prediction of the final expected image quality.

\section*{Acknowledgements}

DES has been funded by the U.S. Department of Energy, the U.S. National Science Foundation, the Ministry of Science and Education of Spain, the Science and Technology Facilities Council (STFC) of Great Britian, the Higher Education Funding Council for England, the National Centre for Supercomputing Applications at the University of Illinois at Urbana-Champaign, the Kavli Institute of Cosmological Physics at the University of Chicago, Financiadora de Estudos e Projetos, Funda\c{c}\~{a}o Carlos ChagasFilho de Amparo \'{a} Pesquisa do Estado do Rio de Janerio, Conselho Nacional de Desenvolvimento Cient\'{i}fico e Tecnol\'{o}gico and the Minist\'{e}rio da Ci\^{e}ncia e Tecnologia and the Collaborating Institutions in DES.

These Collaborating Institutions are Argonne National Laboratories, the University of Cambridge, Centro de Investigaciones Energeticas, Medioambientales y Technologicas-Madrid, the University of Chicago, University College London, DES\-Brazil, Fermilab, the University of Edinburgh, the University of Illinois at Urbana\-Champaign, the Institut de Ciencies de l'Espai (IEEC/CSIC), the Institut de Fisica d'Altes Energies, the Lawrence Berkeley NAtional Laboratory, the University of Michigan, the National Optical Astronomy Observatory, the Ohio State University, the University of Pennsylvania, the University of Portsmouth and the University of Sussex.

We thank STFC for a major grant towards building the DES optical corrector.
DB acknowledges support from an RCUK Academic Fellowship. SB acknowledges support from the Royal Society in the form of a University Research Fellowship and from the European Research Council in the form of a Starting Grant with number 240672. OL acknowledges a Royal Society Wolfson Research Merit Award and Leverhulme Senior Fellowship.

We are grateful to Barnaby Rowe, David Finley, Adam Hawken, Stacey-Jo Dyas and the DES Weak Lensing Working Group for helpful discussions. 

This document has Fermilab ID number FERMILAB-PUB-12-101-AE.

\bibliographystyle{mn2e}
\bibliography{References}

\label{lastpage}

\end{document}